# Stronger when wet: Water-resistant chitinous objects via zero-waste coordination with metal ions

Akshayakumar Kompa, Javier G Fernandez

## Abstract
Plastics have become integral to our society due to their durability and water stability, which is achieved through strong intermolecular interactions. However, these properties also make them persistent disruptors of ecological cycles, in contrast with biological structures, which work with their environments to achieve both excellent mechanical properties and ecological integration. This study takes inspiration from the arthropod cuticle to adapt Earth's second-most abundant organic molecule for use in water. The process involves the vitrification of chitosan with small traces of nickel to create a dynamic network of intermolecular bonds using environmental water, resulting in a material that increases its strength to values well above commodity plastics when wet. The approach preserves the molecule's original chemistry—and therefore its seamless integration into Earth's metabolism—while avoiding the use of the strong organic solvents typically associated with biomolecules. The method demonstrates the potential for a paradigm shift in manufacturing, with zero-waste production of both consumables and large objects that could meet the global demand for plastic.

## Main

The family of materials commonly known as plastics is a fundamental part of modern society. In the decades following the creation of the first plastics, less than a century ago, they rapidly displaced glass, paper, and metal in traditional applications and enabled new ones [1]. With plastics, the economy and habits of society rapidly changed, fueling the mass production of cheap and short-life products. The magnitude of this profound global change in society is comparable to that which came with the manipulation of stone, bronze, or iron, each of which defined its own technological age, suggesting that we might now be living in the "plastic age" [2]. However, in addition to plastic's many advantages, it brought new challenges. Plastics are the first mass-produced materials that are exogenous to Earth's ecological cycles. Thus, every unrecovered piece of plastic that has ever been disposed of has accumulated in some form in a part of Earth's ecosystem [3], from the deepest point of the ocean to remote islands to the tissues of animals and humans [4, 5], and are becoming an increasingly ubiquitous component of food chains [6, 7].

We are now much more aware of the impact of our actions on our ecosystem and thus on the survival and well-being of our species than we were when plastics were first created just a few decades ago, and there is increasing interest in moving away from current unsustainable practices and toward sustainable production models. However, despite all the new policies and increased awareness, our waste production—the measure of our economic and manufacturing inefficiency—has accelerated [8]; the current approach based on waste reduction and recovery, though necessary, has a limited impact on the 254 million tonnes of persistent plastic materials that reach and remain in oceans and landfills every year [9, 10]. A new manufacturing paradigm based on biomaterials that can be reassimilated into the ecological cycles of the Earth without artificial recovery is therefore critical to achieving a sustainable civilization [11].

However, an essential enabling characteristic of plastics is their stability and persistence in water-based environments, which is strongly linked to their biodegradability—or rather, the lack thereof. Plastic materials are made suitable for manufacturing by increasing their crystallinity, crosslinking density, and molecular weights, simultaneously providing water stability and the mechanical characteristics to form standalone structures [12, 13]. However, water plays a pivotal role in the metabolic processes that enable biodegradation [14], and the stability of plastics in the presence of water therefore comes at the expense of compostability. The result is materials that, even when they have bio-based origins, can only be biodegraded in specialized facilities—if at all—making their end of life as bad or worse than every other type of persistent synthetic polymer due to their limited recyclability [15, 16].

That the cost of adapting natural molecules to the plastic-age paradigm is the loss of their integration with Earth's ecological cycles is not because of any limitation of the molecules themselves. Biological systems evolve to use their environment, not isolate themselves from it, producing outstanding structures that both develop and perform in water-rich environments by incorporating water as a central element in their designs [17]. The production of the chitinous cuticle of arthropods is an excellent example of this; it is secreted in a gel-like form into water, hardens to form solid exoskeletons for insects and crustaceans, and performs in humid or underwater environments throughout the animal's life [18]. The sclerotization process, in which the cuticle transitions from a soft hydrogel to an extremely hard structure, is a complex amalgamation of intertwined processes that involve water transport, molecular reorganizations, and mechanical forces [19]. Special attention has recently been given to the role that transition metals (e.g., Zn, Cu, Ni) have in this process and the particularities of the complex relationships they have with the organic components of the cuticle (refer to [20] for an insightful and extensive review of the use of metal ions in biological structures). The natural availability of some of these structural molecules—specifically chitinous polymers—is orders of magnitude greater than the world's demand for plastics, suggesting that they might be good candidates for sustainable production. However, their artificial assembly into solid materials does not recapitulate the creation process and principles that they follow in natural systems [21]. As a result, despite the identical chemistries of the naturally and artificially assembled materials, their differing molecular organizations create completely distinct properties. Specifically, when these materials are artificially assembled without the use of strong organic solvents, they inevitably exhibit high susceptibility to water, requiring the use of coatings before they can be used in product manufacturing [22].

Here, we report the first constructs created with an unmodified structural biopolymer that is formed in a water-based solvent and is particularly suitable for use in contact with water. The constructs are made of nickel-doped chitosan, which incorporates surrounding water into its intermolecular structure to create a network of weak, but highly dynamic, water-mediated bonds, avoiding failure through internal reconfiguration. The result is a material that almost doubles its strength in contact with water, achieving capacities well beyond those of commodity plastics. This is particularly significant both because it is the first time such an outcome has been achieved with a biological material and because chitinous polymers are the second-most abundant organic molecules on Earth—surpassed only by cellulose—allowing the proposed technology to scale-up to a level that will have a global impact. This result is also particularly timely because chitinous polymers are becoming central to the development of sustainable and regional manufacturing processes through their role in valorizing organic waste (e.g., food waste) and the local production of nutrients (**Figure 1a**) [23]. Furthermore, the extensive and efficient production of chitinous polymers in every ecological cycle—particularly by organisms that both produce food and decompose waste—suggests they will even play a key role in creating the efficient, self-sustaining human communities that will allow humanity to colonize other planets [24].

This study was inspired by the serendipitous observation that when zinc is removed from the fangs of the sandworm *Nereis Virens*, they become susceptible to hydration, softening when immersed in water [25]. While most studies on the role of metals in animals' cuticles center on the protein–metal interaction—specifically, the role of histidine—it is already known that the chitinous polymers that form the organic matrix in most cuticles also interact with metals. Indeed, one of the primary uses of chitinous waste from shrimp and crab processing is as a heavy metal flocculant in water treatment systems [26]. While the role of metals in the cuticles of arthropods has largely focused on enhancing their mechanical characteristics, we hypothesized that transition metals might play an essential role in controlling water interactions in chitinous materials. We used nickel specifically because, despite not being as common as other transition metals (e.g., zinc) in the cuticle, it is a ubiquitous micronutrient necessary for life, is water-soluble, and has shown ample versatility in interacting with chitinous polymers in theoretical models [27, 28].

We started our exploration by entrapping different amounts of nickel in a chitinous structure. Chitosan extracted from discarded shrimp (*Penaeus monodon*) shells was dispersed at a 3% concentration in a weak solution of acetic acid at the anaerobic limit (1% acetic acid in water; for comparison, table vinegar ranges from 5 to 8%), to which we added nickel chloride dissolved in water at concentrations from 0.6 to 1.4 M. The water was then evaporated, forcing the vitrification of the polymer into a solid film. The first noticeable change from the presence of nickel in the chitinous films was a green color (λ = 520 nm; films without nickel are yellowish), which is characteristic of nickel (II) compounds. This color became more intense as the concentration increased (**Figure 1b**, **Supplementary Figure 1**).

Theoretically, and if no factor other than the nickel itself is considered, the most stable location for nickel ions in chitosan polymer chains is the space between the primary and secondary hydroxyl groups of consecutive pyranose rings (Position I in **Figure 1c**), where it weakly interacts with the fully coordinated oxygen atom in the ring [27]. When external elements are also considered, the most stable position is between the primary amino and secondary hydroxyl groups (Position II in **Figure 1c**), where the nickel ions can coordinate with water molecules and the sterically available groups of adjacent chitinous chains. It has also been theorized that nickel could take the position between the primary amino and the primary hydroxyl groups of consecutive rings [29]. Fourier-transform infrared (FTIR; **Figure 1d**) spectra of the films show a blue shift of the amide II band from 1542 to 1561 $cm^{-1}$ due to the presence of nickel in the chitinous structure, consistent with interaction with the primary amines (Position II in **Figure 1c**). However, a similar blue shift is also apparent in the band located at 1325 $cm^{-1}$, corresponding to the bending of the -$CH_2$ group, consistent with nickel in the inter-ring position (Position I in **Figure 1c**). The direct contribution of nickel to the spectrum can be observed by the moderate modifications of the 500–700 $cm^{-1}$ region, where the vibrations of the bond between the nickel ions and the hydroxyl groups appear. However, the impact of nickel on the chitosan spectra is not primarily its direct interactions, but rather the changes introduced by the new water molecules that are associated with the nickel ions.

Nickel forms weak hydrogen bonds with water molecules—up to six when in solution—and while the new interactions with the chitinous chains replace some of those molecules, the nickel-doped films incorporate several times as many water molecules as the number of nickel ions introduced. The effect of this additional water is observable in the band at 1628 cm$^{-1}$, which corresponds to the bending vibration of the hydroxyl groups in water molecules. While this vibration is overshadowed by the amide I band (1636 cm$^{-1}$) in pristine chitosan films, it completely dominates the fingerprint region in nickel-doped films with even the smallest amount of nickel and rapidly grows as the concentration increases. A similar effect can also be clearly seen in the intensity of the broad band at 3250 cm$^{-1}$ (**Figure 1e**), where the intensity of the vibrations corresponding to the stretching of the O-H bond rapidly increases with the amount of nickel in the system.

The additional water also strongly impacts the crystal structure of the chitosan films (**Figure 1f**). Regular chitosan films have a hydrated crystal structure, in which water molecules mediate many of the intermolecular interactions. It has recently been demonstrated that strain stiffening, which reorganizes chitosan films into a more crystalline structure using external forces, also results in closer packing by reducing the free volume and expelling water molecules from the material as new chain–chain direct interactions replace their binding sites [30]. A similar but opposite effect can be observed in the nickel-doped films, as the additional nickel and its associated water results in lower crystallinity. Considering only the changes in the chitinous structure, the inclusion of nickel and its associated water into the chitinous structure might be (wrongly) seen as plasticization; indeed, a common approach to increasing the flexibility or manufacturability of long polymers is to disrupt their intermolecular bonds, although this occurs at the cost of decreasing their tensile strength. However, in the nickel-doped films, the disruption of the intermolecular bonds by the additional water and the lower crystallinity does not have the expected negative impact on their mechanical characteristics.

All the samples—independent of their nickel content and therefore their crystallinity—have tensile strengths in the range of 30 to 40 Mpa (**Figure 1g**), which is similar to commodity plastics. Despite the additional water and lower crystallinity, the use of low concentrations (less than 0.8 M) of nickel has a minimal impact on the mechanical properties of the films, with insignificant changes in their strengths or elasticity with respect to pure chitosan films. Beyond 1 M concentrations, however, the strength of the material is preserved while its Young's modulus—a measure of stiffness—falls significantly, marking the material's increased ability to be stretched and, therefore, to absorb more energy before breaking—in other words, greater toughness. Introducing nickel into the chitosan structure therefore increases both the material's flexibility as a plasticizer and its strength as a crosslinker, which are usually considered incompatible properties. This result cannot be overstated: the ability of a material to be both tough and strong simultaneously has been a chimeric goal in the field of structural materials that, as a feature unique to biological composites, has been the leading motivation for research into bioinspired materials [31]. Furthermore, the ability to tune the properties of a biopolymer to different mechanical characteristics without sacrificing strength enables a single material and production process to be adapted to different functions. In biological systems, this phenomenon makes possible, for example, the insect cuticle, which is a continuous and multifunctional structure made of very stiff parts (e.g., wings) and very elastic parts (e.g., joints) with only minor compositional changes. This use in nature of few, but very versatile, materials is the result of strong evolutionary pressure toward efficiency—the so-called survival of the cheapest [32].

This principle is directly applicable to product design, optimizing both cost and environmental impact by minimizing the number of materials used, simplifying both production and end-of-life management.

While the creation of an unmodified biological material with tunable hardness and constant strength is already an exceptional result, the inclusion of nickel ions in the chitinous matrix provides an even more extraordinary effect: the material gets stronger when it is immersed in water (**Figure 1h**, **Supplementary Video 1**). The strengths of all the nickel-doped films, except the one with the lowest concentration of nickel, remained constant or increased when immersed in water. This is particularly notable in the case of samples made with a 0.8 M concentration of nickel, which had a strength increase of almost 50% upon immersion. We hypothesized that there is an optimal balance in the roles of the nickel and water in simultaneously enhancing the intermolecular bonds through new interactions and disrupting them by preventing direct chain–chain interactions. We therefore focused the rest of the study on the material produced using a 0.8 M concentration of nickel.

The nickel-doped films, which in dry conditions have a tensile strength of 36.12 ± 2.21 MPa—within the strength range of commodity plastics (e.g., polypropylene, polystyrene, polylactic acid)—show an increased tensile strength of 53.01 ± 1.68 MPa when immersed in water, placing them in the range of engineering plastics (e.g., polycarbonate, polyethylene terephthalate glycol, polyoxymethylene; **Figure 2a**). Using a 0.8 M concentration for fabricating the nickel-doped films appears optimal for achieving this mechanical improvement in water but at the cost of also incorporating significant amounts of nickel that is irrelevant to that enhancement. This can be observed at the macroscopic level as the leaching of nickel and the behavior of the nickel-doped films when first immersed in water: freshly made films show a permanent mechanical change with respect to their initial state after the first immersion. Subsequently, nickel-doped films that are successively hydrated and then dried at 60°C for 24 h transition between two mechanically distinct states, both different from that of a freshly made film (**Figure 2b**). This macroscopic change after the first immersion is also observable in the accumulation of nickel on the surface of a freshly made film, which later disappears (**Figure 2c**, **Supplementary Figure 2**).

At the molecular level, the C1s x-ray photoelectron spectrum of the surface of a nickel-doped film shows that three peaks (C-(C, H) (aliphatic), C-(N, O), and C=O/O-C-C) have higher binding energies than the same peaks in a pure chitosan film (**Figure 2d**, **Supplementary Table 1**). There is also a notable reduction of the area under the deconvoluted C-(N, O) peak (i.e., 286.37 eV), confirming that the main interaction between nickel and chitosan occurs through the latter's side functional groups [27, 29]. However, after the first immersion of a nickel-doped film in water, those differences disappear, resulting in a surface analysis that resembles that of pure chitosan films. The same effect is observed in the O1s and N1s spectra (**Supplementary Figure 3**), supporting the hypothesis that the permanent change in the material is because the loosely bound nickel and associated water is removed upon first immersion. This hypothesis is also substantiated by the weight loss of the nickel-doped films immediately after immersion, in contrast to the weight gain normally expected of a typically hydrophilic material (e.g., chitosan) due to water uptake (**Supplementary Figure 4**).

In nature, chitinous materials are organized as hydrated crystals in which the relatively long-range interactions between the functional groups of adjacent polymer chains are enabled by intermediate water molecules [33]. The inclusion of nickel during the formation of the chitinous structure brings a new level of dynamism to the material. Nickel can form weak hydrogen bonds with multiple water molecules— whether dissociated or not—and with the oxygen and nitrogen in the functional groups of the chitosan chains. While some water is bound within the polymeric chains, large amounts are also introduced from the environment alongside the nickel (**Figure 2e**). The result is a structure of stiff polymeric chains that are bound together by a combination of direct interchain bonds and weak and rapidly configurable bonds, which are mediated by highly motile particles (i.e., nickel and water) trapped within the structure (**Supplementary Figures 5** and **6**). This versatility to bond in multiple ways also includes combinations that do not contribute to the structure, such as nickel and water molecules without significant bonds to the polymeric chains. We hypothesize that these nickel particles and associated water molecules that do not contribute to the structure are the ones released during the first immersion. This hypothesis is consistent with the observation that 43% of a freshly made nickel-doped film's weight is water, but this drops to 20% after the first immersion, which is still significantly higher than the weight of water in pure chitosan films (17%; **Figure 2f**). Similarly, even while the films keep their strength underwater, 87.18 ± 2.72% of the entrapped nickel is weakly bound to the chitosan and is released during the first immersion (**Figure 2g**). This indicates that, despite the amount of nickel introduced, only one nickel ion for every 7.91 pyran rings is needed to produce the water-strengthening effect in chitosan. In a macroscopic context, this means that the nickel content of a discarded AAA battery (2.2 g) would be sufficient to manufacture more than a dozen typical plastic cups (4.7 g each) using nickel-doped chitosan.

The release of the inconsequential nickel during the first immersion could be seen as an optimization of the material, in which the arbitrary entrapment of the nickel needed to saturate the system is followed by a process that removes all but the nickel that contributes to the material's structure. This process is done in water, which also happens to be the environment in which the films are formed. The resulting "waste" from the optimization process is therefore itself an ingredient (i.e., nickel and water) for producing the films. With this in mind, we developed a production cycle in which the water used to remove the inconsequential nickel is used as an input for fabricating films (**Figure 3a**). Using this zero-waste process, we produced several objects with an approach similar to the use of chitosan in product manufacturing—that is, vitrifying it on a positive mold [22]. This approach enabled the production of common plasticware, such as containers and cups (**Figure 3b**). Note that the purpose of these objects was to explore the material's malleability rather than its ability to replace products that might no longer be justifiable in a society that can use biological materials and regionalized production. Nevertheless, the nickel-doped chitosan containers showed that the biological molecule not only gains strength in water environments but also can contain water as effectively as common plastics, demonstrating the material's suitability for such tasks (**Figure 3c**, **Supplementary Video 2**). We extended the studies to producing negative replicas, which is a non-trivial task because the vitrification process entails a shrinking of the object as it forms, preventing the use of molds unless they maintain contact during the process. To overcome this limitation, we fabricated a random positioning machine (a two-axis clinostat; **Figure 3d**, **Supplementary Video 3**) with two independently perpendicular frames and an attached negative mold. The continuous repositioning of the mold forced the polymeric solution to maintain contact during the vitrification, enabling the molding of closed geometries and a qualitative improvement in the aesthetics of the objects with respect to those produced with positive molds (**Figure 3e**).

In the last few decades, the materials proposed as ecologically integrated plastics have focused on recovery processes [34] or uncommon biological components with limited scalability [35]. The approach here is based on the biomimetic coordination of a transitional metal with chitin, Earth's second-most abundant organic material with an estimated renewable production rate of $10^{11}$ tonnes per year [36]—the equivalent of *three centuries* of plastic production. Such production is sustainable, seamlessly integrated into Earth's ecological cycles (**Supplementary Figure 8**), and easily reproducible in urban environments through the bioconversion of food and other organic wastes [23]. This abundance enables the scaling of the results presented here to levels unprecedented for a new technology but, more importantly, the theoretical upscaling of production to quantities with a global impact and the regionalization of that production. To explore the potential upscaling—even within the limited confines of a research laboratory—we produced a 1m$^2$ chitin-doped film and tested its ability to hold weight after 24h of water immersion (**Supplementary Video 4**). Despite the unparalleled scaling of such a newly incepted material, we then went further with the production of a film three times larger without problems (**Figure 3f**, **Supplementary Video 4)**, demonstrating an absence of scalability restrictions and highlighting the potential for rapid scaling of the results and principles presented here to ecologically relevant scales.

In conclusion, we have demonstrated that the interaction between transition metals—specifically nickel—and chitosan creates unique material properties for the molecule in its native form, from the tunability of elastic properties without impacting material strength to improving mechanical performance underwater. Unlike the usual approach of fitting biological molecules into the synthetic plastic paradigm, the production process presented here applies the principles of bioinspired manufacturing by adapting biological strategies to use unmodified biological components, applying water-based and environmentally friendly chemistries, and minimizing waste production [37, 38]. This strategy preserves the seamless integration of the chitinous polymer into Earth's ecological cycles without relying on human intervention or recovery.

The results presented here represent a new paradigm for manufacturing in a biological environment and a departure from the current approach that assumes inertness to be unavoidably associated with resilience. In the case of polymer-based materials, resilience is achieved through heavily crosslinked long molecules that depend on strong and exhaustive internal links to produce materials that are unresponsive to the surrounding environment. Here, resilience is achieved in a biomimetic way, using the environment instead of fighting it, relying on water transport between the nickel-doped chitosan and the environment to create a dynamic structure of weak and long-range intermolecular bonds in continuous reconfiguration.

Because of the outstanding mechanical properties of the nickel-doped chitinous materials in water-based environments, the absence of a human immune response, and existing FDA approval for medical uses of both nickel [39] and chitosan [40] individually, we foresee immediate applications of these results in the medical field and as a water-proof coating for biomaterials [37] while the proposed shaping technologies are perfected and upscaled for general use. However, in a world that generates 400 million tonnes of plastic every year, much of it specifically for its performance in water environments [41], we believe the approach presented here and the principles and materials upon which it is based have implications at an unprecedented scale. They mark a paradigm shift toward bioinspired manufacturing based on ubiquitous biological materials, regional production, and ecological integration, which can address the limitations of current manufacturing and curb humanity's persistent waste production.

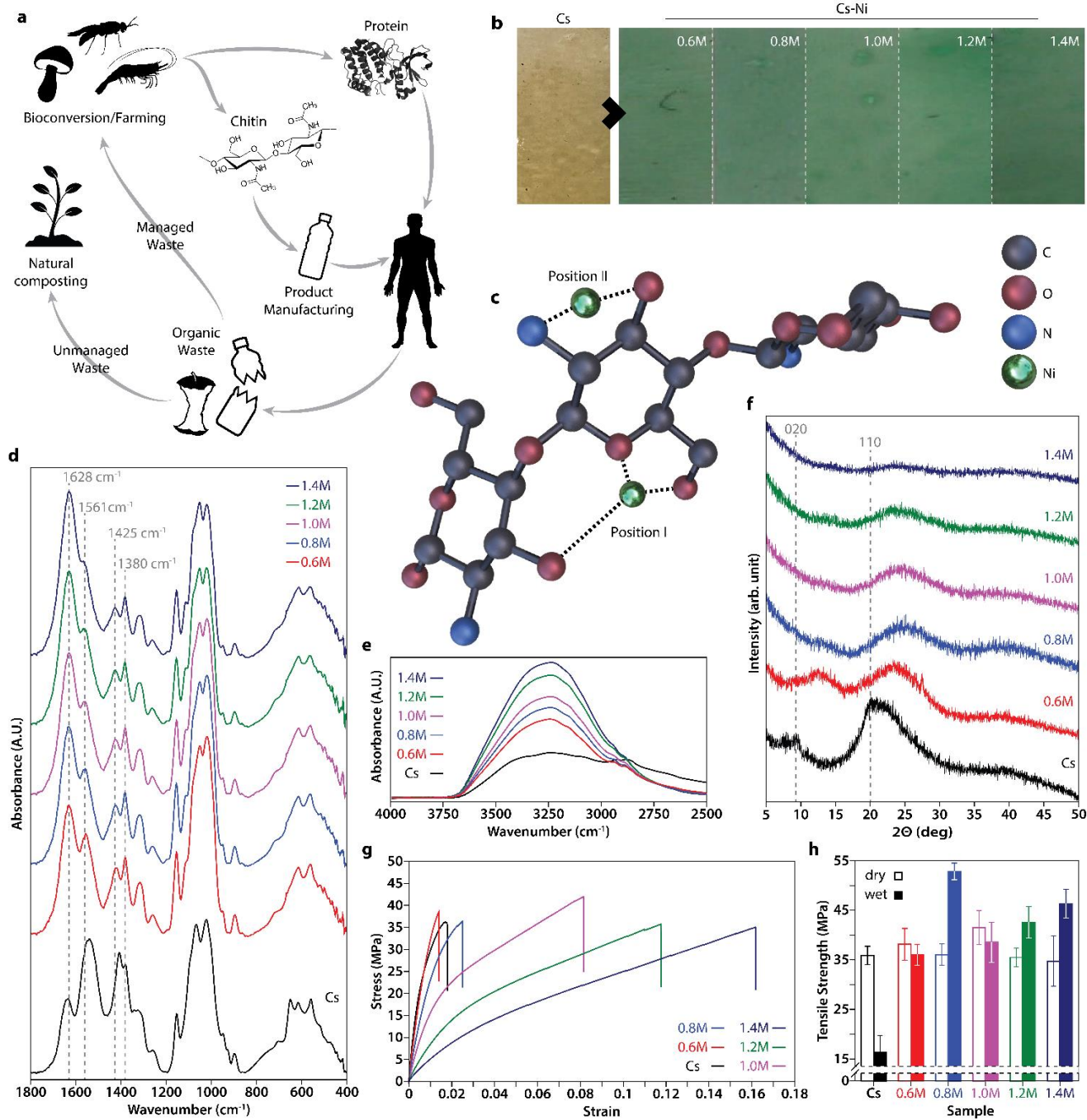

**Fig. 1 – Production of nickel-doped chitosan**. **a**. Local production of chitinous polymers as part of regionalized circular economies. Chitin and chitosan, typically a byproduct of the shrimp and crab processing industry, are structural components in most heterotrophs used in the bioconversion of organic waste and the local production of nutrients. As part of any ecological cycle, chitin can be reincorporated into circular production cycles through waste management or through natural ecological cycles if unmanaged. **b**. Qualitative comparison of the color of the vitrified chitinous films with various amounts of nickel trapped in them (refer to Supplementary Figure 1 for a quantitative analysis). **c**. Most likely locations of nickel ions within a chitosan chain. The different locations result from different assumptions about the degree of deacetylation, water content, and crystal structure of the polymer, which vary across the material. In reality, the distribution of the nickel involves all these options and their variations.

**d**. Infrared Fourier-transform spectra of nickel-doped chitosan films with different amounts of nickel. The differences between the spectra resulting from the nickel ions are overshadowed by the changes caused by the new water molecules introduced into the material in coordination with the nickel ions and the sterically available side groups in the chitosan chains. **e**. Chitinous films' spectra in the region dominated by water's O-H bond vibrations and normalized to the carbohydrate skeletal vibrations (800–500 cm$^{-1}$, inherent to the chitosan structure), showing the increasing nickel-driven introduction of intermolecular water in the chitinous films. **f**. Crystal conformation of chitinous films. In the absence of nickel, vitrified chitosan films show two distinct peaks typical to their hydrated crystal conformation at 9.5° and 20° and a broad, amorphous region in the 15–30° range. As the amount of nickel and water increases, the amorphous regions come to dominate the structure. **g**. Representative examples of the tensile stress–strain curves of the nickel-doped films. Concentrations of nickel below 0.8 M have a limited impact on the material's tensile strength. Interestingly, beyond that point, the material gains elasticity without sacrificing strength, simultaneously achieving strength and toughness, which is characteristic of the functional versatility of structural biomaterials. **h**. The tensile strength of films with different nickel contents before (empty bars) and after (solid bars) immersion in water. Color codes are consistent for all panels, with Cs (black) representing pristine chitosan films (without nickel) and other colors representing the different concentrations of nickel in the initial solution, from 0.6 M to 1.4 M.

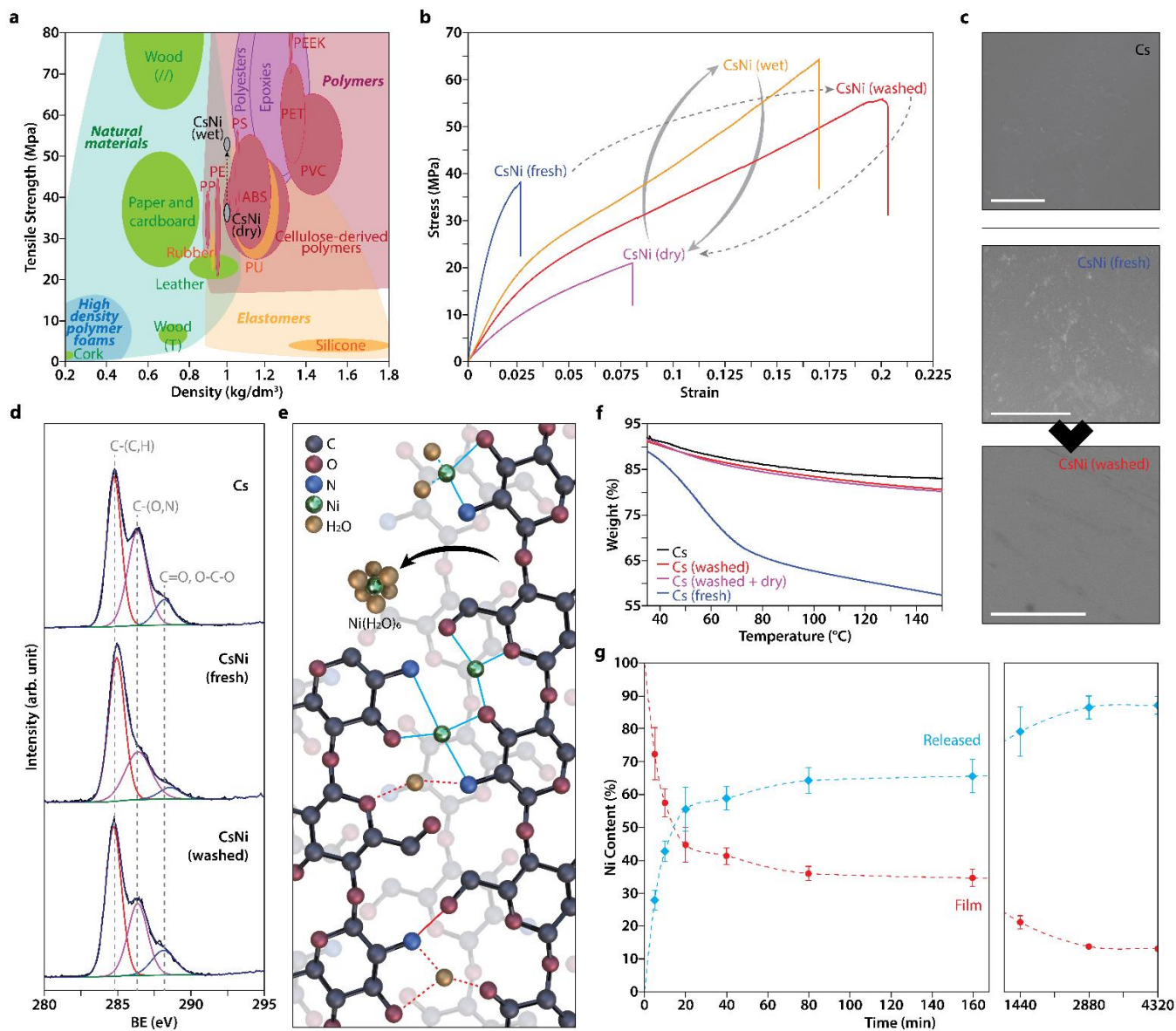

**Fig. 2 – Nickel-doped chitosan in water**. **a**. Ashby plot representing tensile strength with respect to density for natural and synthetic materials. Nickel-doped chitosan in a water environment (labeled "CsNi wet") or after removing the additional water at 60°C for 24 h ("CsNi dry") are also represented. **b**. Representative mechanical curves for the permanent changes of the nickel-doped chitosan after the release of inconsequential nickel ions ("washed") and the reversible changes from extracting intermolecular water by oven-drying ("dry") and replenishing it by wetting the films ("wet"). **c**. Scanning electron microscopic images of the films' surfaces. Cs (top) is a pure (i.e., without nickel) chitosan film, and CsNi (fresh) is a nickel-doped film immediately after fabrication. The white precipitates in the image are nickel accumulations on the surface. The excess nickel on the surface disappears after the first immersion ("CsNi washed"). **d**. C1s x-ray photoelectron spectra of a chitosan film without nickel (top) and a nickel-doped film before (middle) and after (bottom) first immersion in water. In the latter, the interactions greatly resemble those of pristine films. **e**. Molecular diagram of the role of nickel and water in vitrified chitinous films. Interchain interactions are possible directly (solid lines) and through water molecules (dashed lines). The inclusion of nickel, which can simultaneously link several chains and water molecules, adds a new level of possible weak and easily reconfigurable bonds (blue lines). Among the possibilities is the complete or partial coordination of nickel ions with water, without significant bonds to the polymeric chains, and their release to the environment. **f**. Thermogravimetric analysis of the chitinous films in the temperature range for water evaporation; 43% of a freshly made nickel-doped chitosan film (blue line) is water. This percentage is reduced to 20% when the films are immersed in water for the first time (purple line), and it remains stable after drying for 24 h at 60°C (red line) (see Supplementary Figure 4 for an analysis of the weight loss during the first immersion). Notably, this amount of water is significantly higher than that in regular chitosan films (i.e., without nickel, black line). **g**. Colorimetric analysis of the release of inconsequential nickel from the chitinous film to its surroundings during its first contact with water. Overall, only 13% of the nickel introduced contributes to the intermolecular bonding of the material, which is equivalent to about one nickel atom per eight saccharide rings, while the remaining 87% is released to the surrounding water. Most of the inconsequential nickel is released in minutes (left side of the graph). After 24 h, no more measurable nickel is being released from the films (right-side graph).

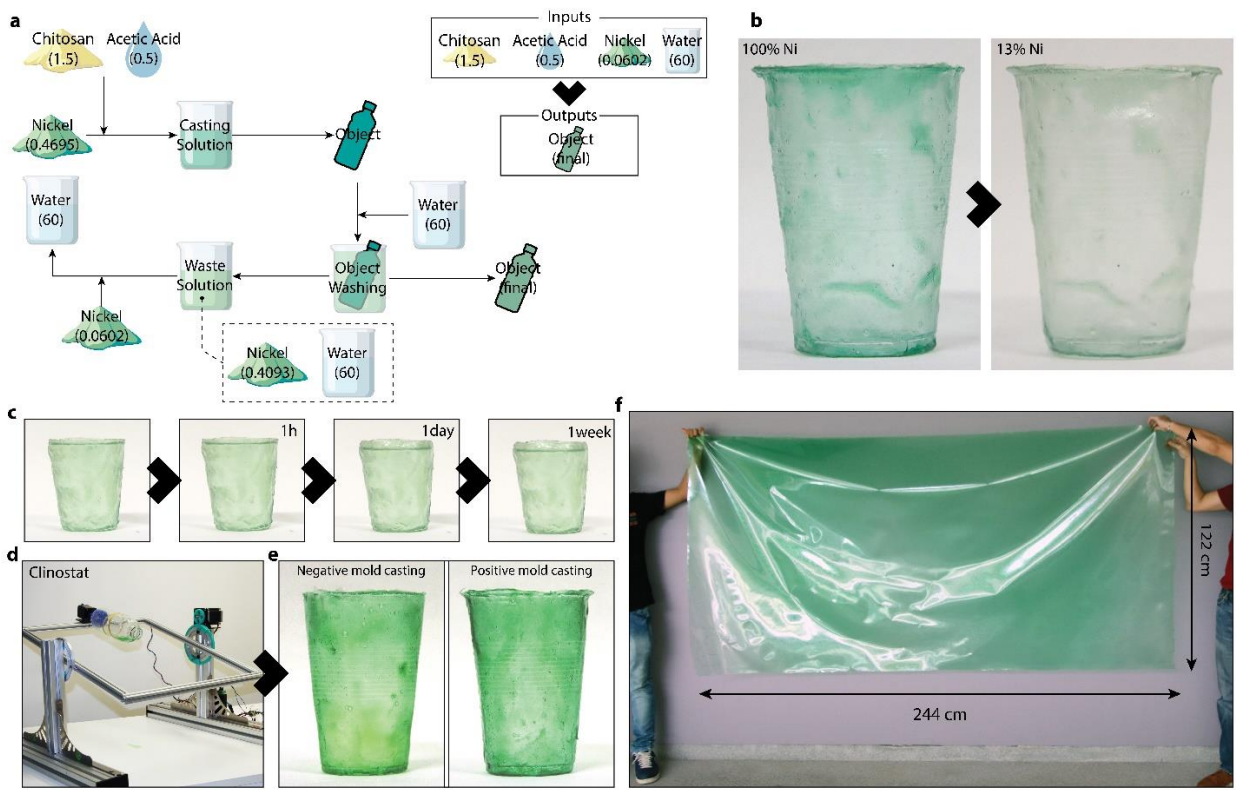

**Fig. 3 – Product manufacturing**. **a**. Diagram of zero-waste production of chitinous objects. The inconsequential nickel released during the optimization of one object is used as a primary component for the next one by topping up the nickel. The system ensures 100% utilization of nickel despite the need to saturate the material with large amounts of it during vitrification to achieve the water-strengthening effect (see Figure 1h). The top-left corner shows the actual inputs and outputs of the process by weight. As a reference, the drinking cup in the next panel weighs 4.7 g. **b**. Replica of a drinking cup made of nickel-doped chitosan formed using a negative mold (i.e., the material vitrifies against the mold's outside walls). On the right is the same object after the inconsequential nickel has been removed by immersion in water—a process that has no apparent impact on its geometry. See **Supplementary Figure 7** for closer images of the color change due to the release of the inconsequential nickel. **c**. Images of a nickel-doped chitosan cup filled with water to demonstrate the impermeability of the material. A detailed recording of the first 24 h is in Supplementary Video 2. **d**. Picture of the clinostat used to replicate negative molds for the nickel-doped chitosan. The continuous movement forces the material to vitrify against the inner walls, enabling the fabrication of closed geometries and much more accurate and high-quality objects (see Supplementary Video 3 for more detailed information). **e**. Comparison of the same objects fabricated using negative (i.e., with the clinostat in the previous panel) and positive molds. **f**. An example of a three-square-meter film made of nickel-doped chitosan that was used to test the scalability of the process. Detailed information on this construct can be found in Supplementary Video 4.

## Supplementary Information

The supplementary information comprises materials and methods, eight additional figures, and four videos.

## Acknowledgments

The authors thank Dr. Cedric Finet from the National University of Singapore (NUS), the Institute of Materials Research and Engineering (IMRE), Agency for Science, Technology and Research (ASTAR) for the chemical analysis. We want to thank Dr. Xueliang Li for his assistance with the XRD measurements and Ms. Sarah Wasifa Ferdousi for her help building the clinostat.